# Highly tunable ferroelectricity in hybrid improper ferroelectric $Sr_3Sn_2O_7$

*Xianghan Xu, Yazhong Wang, Fei-Ting Huang, Kai Du, Elizabeth A. Nowadnick and Sang-Wook Cheong*[*]

Xianghan Xu, Yazhong Wang, Fei-Ting Huang, Kai Du, and Sang-Wook Cheong
Rutgers Center for Emergent Materials and Department of Physics and Astronomy,
Rutgers University, Piscataway, NJ 08854, USA
E-mail: sangc@physics.rutgers.edu

Elizabeth A. Nowadnick
Department of Materials Science and Engineering, University of California, Merced, Merced, CA 95343, USA

**Abstract**

The successful theoretical prediction and experimental demonstration of hybrid improper ferroelectricity (HIF) provides a new pathway to couple octahedral rotations, ferroelectricity, and magnetism in complex materials. To enable technological applications, a HIF with a small coercive field is desirable. We successfully grow $Sr_3Sn_2O_7$ single crystals, and discover that they exhibit the smallest electric coercive field at room temperature among all known HIFs. Furthermore, we demonstate that a small external stress can repeatedly erase and re-generate ferroelastic domains. In addition, using in-plane piezo-response force microscopy, we charaterize abundant charged and neutral domain walls. The observed small electrical and mechanical coercive field values are in accordance with the results of our first-principles calculations on $Sr_3Sn_2O_7$, which show low energy barriers for both 90 ° and 180 ° polarization switching compared to those in other experimentally demonstrated HIFs. Our findings represent an advance towards the possible technological implemetation of functional HIFs.

## 1. Introduction

Hybrid improper ferroelectrics (HIFs) are materials with spontaneous electric polarization that couples to multiple non-polar lattice distortions— a new type of ferroelectricity that was only discovered and appreciated over the last two decades.[1] [2] [3] Hybrid improper ferroelectricity was first realized experimentally in 2008 in $PbTiO_3/SrTiO_3$ multilayers[1] and more recently in 2015 experimental switching of electric polarization in $Ca_3Ti_2O_7$ bulk crystals was demonstrated.[3] Over the past few years, a number of HIF candidates have been identified using a combination of group-theoretical symmetry analysis and first-principle calculations,[4] [5] and HIF has been revealed in Ruddlesden-Popper (RP) phases,[3] [6] [7] Dion-Jacobson layered perovskite oxides,[8] [9] perovskite superlattices [1] [10] [11] as well as metal-organic molecular perovskites.[12] [13] The potential to cross-couple order parameters within HIFs makes them of great interest for materials by design, for example to realize room-temperature magnetoelectricity[14] or large uniaxial negative thermal expansion.[15] However, the large switching barrier associated with the geometric origin of HIF still hinders the development of functional HIFs.[16]

Recent studies have found that ferroelastic domain walls in orthorhombic HIFs mediate a 90° rotation of the polarization vector and are mobile, switchable,[17] [18] [19] [20] and incorporated as Néel-type domain walls screening head-to-head and tail-to-tail dipoles in 180° domain walls.[5] Due to these factors, the predicted inaccessibly large switching barriers of HIFs turned out to be accessible experimentally, partially due to a multi-step switching mechanism.[3] [20] [18] For example, the ferroelectric polarization switching coercive field is 120-180 kV cm$^{-1}$ in $Ca_{3-x}Sr_xTi_2O_7$ (x=0, 0.54, 0.85) single crystal,[3] 75-160 kV cm$^{-1}$ in $Ca_3(Ti_{1-x}Mn_x)_2O_7$ ceramic[21] and single crystal,[22] 150 kV cm$^{-1}$ in $Sr_3Sn_2O_7$ polycrystalline sample,[18] 100 kV cm$^{-1}$ in $Sr_3Zr_2O_7$ polycrystalline sample[6] and 180 kV cm$^{-1}$ in $CsBiNb_2O_7$[23] polycrystalline sample. Nevertheless, these electric coercive fields are an order of magnitude larger than that those of proper ferroelectrics such as $Pb(Zr,Ti)O_3$[24] and $BiFeO_3$,[25] which have coercive fields of 7 kV cm$^{-1}$ and 15 kV cm$^{-1}$, respectively.

It is well known that the Goldschmidt tolerance factor $t$ can predict octahedral rotation amplitudes in perovskites based on the ionic size mismatch.[26] A linear relationship between the tolerance factor $t$ and the Curie temperature ($T_c$) of RP-type HIFs has been empirically revealed,[27] implying a possible predictive relationship between the amplitude of oxygen octahedral rotations and the ferroelectric switching barrier. Specifically, the switching barriers may reduce with increasing tolerance factor. However, no HIF with polarization switching barrier comparable to proper ferroelectrics such as $Pb(Zr,Ti)O_3$ and $BiFeO_3$ has been realized experimentally, even in materials with large tolerance factors. We notice that most works so far presented poly-crystalline samples, in which electric leakage and pinning effects at grain boundaries could impede the polarization switching process. Part of the reason for this is that the preparation of high-quality decently sized single crystals of RP-type HIFs is challenging.

Here, we report the first successful growth of $Sr_3Sn_2O_7$ single crystals, which have the largest tolerance factor of all experimentally reported HIFs to date. We demonstrate that $Sr_3Sn_2O_7$ single crystals have the smallest polarization switching coercive field, 39 kV cm$^{-1}$ (47 kV cm$^{-1}$) along the [100] ([110]) direction, among all experimentally discovered bulk HIFs. In addition, we reveal that $Sr_3Sn_2O_7$ is a ferroelectric-ferroelastic crystal, where ferroelectricity and ferroelasticity are strongly coupled, i.e. detwinning can be achieved by electric field and vice versa. In-plane piezo-response force microscope (IP-PFM) images show intriguing ferroelectric-ferroelastic domain structures comprised of abundant meandering charged domain walls. We interpret these experimental results using density functional theory (DFT) calculations of the energetics of 90° and 180° ferroelectric switching paths. Taken together, these results point towards new research opportunities, both in understanding how to design bulk HIFs with small coercive fields, and in the application of erasing and writing charged walls in electronic devices.

## 2. Results
### 2.1 Small electric coercive field in $Sr_3Sn_2O_7$

$Sr_3Sn_2O_7$, $(ABO_3)_nA'O$ RP-phase (n=2) HIF, is the first room temperature ferroelectric (FE) Sn insulator with a large tolerance factor (*t*=0.957) and a low $T_c$ (410 K)[27]. The unusual polarization switching kinetics, where neutral FE domain walls (DWs) move fast while charged FE DWs do not move, have been observed through an in situ poling process using a dark-field transmission electron microscopy technique (DF-TEM)[18]. Furthermore, $Sm^{3+}$-doped $Sr_3Sn_2O_7$ is a mechanoluminescent material in which strain activates light emission.[28] In addition to its mobile charged DWs and strain sensitivity, $Sr_3Sn_2O_7$ is Pb-free, making it an ideal candidate not only for fundamental physics studies but also for environmentally-friendly applications. However, the challenging combination of volatile raw materials and high melting temperature has so far prevented the synthesis of $Sr_3Sn_2O_7$ single crystals.

$Sr_3Sn_2O_7$ single crystals were grown using a laser floating zone furnace, which has a well-focused laser beam on the melting zone and a steep temperature gradient at the interface between the liquid and the solid. This helps overcome the conflict between the high melting temperature (over 2000 °C) and the serious evaporation loss of $SnO_2$ around this temperature. The as-grown $Sr_3Sn_2O_7$ single crystal boule is shown in **Figure 1a** and is confirmed to crystallize in space group $A2_1am$ with lattice parameters $a$=5.733 Å, $b$=5.7057 Å, and $c$=20.6637 Å. The polarization lies along the $a$-axis ($a$>$b$) and comes mainly from Sr displacements, which as shown in **Figures 1b-1c** are induced by a combination of an in-phase $SnO_6$ octahedral rotation about the c-axis ($a^0a^0c^+$) and an out-of-phase octahedral tilting about the orthorhombic $b$-axis ($a^-a^-c^0$). A cleaved thin piece is transparent, consistent with the reported large magnitude (4.13 eV) of the optical gap[28] and shows straight orthorhombic twin walls. Samples are prepared for bulk polarization ($P$) versus electric field ($E$) hysteresis loop measurements along two directions: one is polished so that electric fields can be applied along the [110] direction; the other one is polished 45° with respect to the twin boundaries so that the electric fields can be applied along the [100] ([010] in nearby twin domains) direction.

**Figure 1d** shows the room-temperature $P(E)$ loops measured in oil at frequency $f$=270 Hz using the positive-up-negative-down (PUND) method to get rid of leakage and capacitance contributions. The net remanent electric polarization along the [110] direction is 1.15 μC cm$^{-2}$. If only 180° polarization flipping is considered, the remnant polarization along [100] should be $1/\sqrt{2}$ times that along the [110] direction. However, our measurement shows the remnant polarization along [100] is actually 1.53 μC cm$^{-2}$, which is almost $\sqrt{2}$ times that along [110]. This suggests that both 180° and 90° polarization flipping happens when an external electric field is applied along the [100] direction, which indicates that an electric field can erase ferroelastic DWs. The compensated current density J(E) versus electric field curves are shown in **Figure 1e**. The polarization switching coercive field corresponding to the peak position of the $J(E)$ curve at room temperature is 39 kV cm$^{-1}$ along the [100] direction and 47 kV cm$^{-1}$ along the [110] direction, which is the smallest in all experimentally discovered bulk HIFs to our knowledge. The normalized $P(E)$ loops of the $Sr_3Sn_2O_7$ single crystal (blue curve along [110] and red curve along [100]), $Sr_3Sn_2O_7$ poly (black curve) and the prototypical HIF $Ca_3Ti_2O_7$ single crystal (yellow curve along [110]) are shown in **Figure 1f**. The coercive field of the $Sr_3Sn_2O_7$ single crystal is a factor of three smaller compared to the other materials.

**2.2 Ferroelastic domains under strain**

In addition to a small electric coercive field, we notice that the application of an external strain can easily erase and re-generate the ferroelastic (FA) domains, as shown in **Figure 2**. Figure 2a shows a schematic geometry of our transmission polarized optical microscope (t-POM) measurement. We cleave an $ab$-plane transparent thin plate-like sample and polish it to a square shape with boundaries 45° to the FA DWs. We remove strain induced during the specimen preparation through a post annealing at 1400°C followed by a 50°C per hour cooling. The dark-bright contrast in the t-POM image, as shown in **Figure 2b,** represents the distribution of FA domains in the initial state. We then apply a uniaxial stress along the horizontal direction, i.e. 45 ° with respect to the FA DWs, by squeezing the left and right glass sliders, as indicated by the blue arrows in **Figure 2c**. **Figure 2c** shows the final state t-POM image with a single FA domain after removing the applied horizontal strain. The FA domains with polarization directions parallel to the strain direction appear shrunken and erased, which will be proved by combining with IP-PFM measurements later. Further, the FA domains can be re-written (**Figure 2d**) and expanded (**Figure 2e**) in the presence of strain applied along the perpendicular direction as shown by the vertical pink arrows. **Figures 2f and 2g** illustrate the underlying mechanism. **Figure 2f** shows the *a-b* plane crystallographic unit cells crossing the FA domain walls (dotted black lines). As explained in **Figure 1a**, the Sr displacements define the bulk polarization direction, i.e. the long *a*-axis ($a > b$). Crossing the FA domain walls indicated by dotted black lines, the orthorhombic *a-b* axis rotates by 90 °. When a vertical stress is applied (**Figure 2f-g),** the nearby domain (region A) with the longer *a*-axis parallel to the strain direction will not be favored, accompanying with the expansion of the middle FA domains (region B). When a horizontal stress is applied**,** the nearby domain (region A) will become favored.

Our results described above demonstrate that external strain can erase FA domains with electric polarization parallel to it and re-write FA domains with the electric polarization perpendicular to it. Furthermore, this erasing and re-writing of FA domains can be repeated multiple times. To study the strain induced twinning and detwinning phenomenon quantitatively, we record an in-situ movie (see S.I.) by microscope of a stress-and-release process, and derive a half-strain hysteresis loop, shown in **Figure 3**. For each data point, the major domain percent area is given by the ratio of the major domain area to the total area of images, and the strain is given by $\Delta L/L$, which is the change of sample length divided by the original sample length without stress. Here, two factors contribute to the strain: first, the length change of each domain under stress, and second, the length change from twinning and detwinning since the crystallographic *a* and *b* axes have different lengths. The first factor is non-hysteretic, while the second factor could produce a remnant strain, since after the stress-and-release process, the major domain ratio does not match its original value. This also explains why the strain does not return to zero at the end of the half loop. Our data show that twinning and detwinning can repeatably occur in $Sr_3Sn_2O_7$ crystal under external stress, and that strain can be partially retained even after the stress is released. Thus, $Sr_3Sn_2O_7$ belongs to the family of ferroelectric-ferroelastic materials that includes the well-known Rochelle salt[29] and lead lanthanum zirconate tianate.[30]

**2.3 DFT calculations**

To understand the small electric and stress coercivity of the $Sr_3Sn_2O_7$ single crystal, we calculate the energetics of different polarization switching paths using DFT and the nudged-elastic-band (NEB) method, as shown in **Figure 4.** Along each path, the polarization reverses by 180 ° as the switching coordinate changes from 0 to 1. Due to the HIF mechanism, the polarization is trilinearly coupled to the $a^-a^-c^0$ octahedral tilt and the $a^0a^0c^+$ octahedral rotation, so when the polarization reverses by 180 °, the sense of either the tilt or the rotation also must reverse. As a result, there are several symmetry-distinct pathways to reverse the polarization. There are two 'one-step' paths, where the polarization and the octahedral tilt ($a^-a^-c^0$, blue curve) or the octahedral rotation ($a^0a^0c^+$, green curve) amplitudes go through zero at the midpoint of the path. In addition, there are several symmetry-distinct two-step paths[20]. Here we consider 'two-step' paths where the polarization reverses by making two 90 ° steps (red curve) and where the polarization reverses by passing through a low-energy antipolar structural phase (black curve). The antipolar phase contains the same pattern of Sr displacements in each perovskite slab as in the polar ground state (**Figure 1b**), but the displacements in adjacent slabs are in opposite directions. The energy barriers for the switching paths in **Figure 4** are 125 meV f.u.$^{-1}$ (reverse tilt), 41 meV f.u.$^{-1}$ (reverse rotation), and 39 meV f.u.$^{-1}$ (90 ° switching), and 24 meV f.u.$^{-1}$ (antipolar switching). These barriers are about a factor of two lower than the corresponding barriers for the prototypical HIF $Ca_3Ti_2O_7$ (which are 180 mev f.u.$^{-1}$, 112 mev f.u.$^{-1}$, 82 mev f.u.$^{-1}$, and 64 meV f.u.$^{-1}$, respectively), which may explain why $Sr_3Sn_2O_7$ has a lower coercive field. In addition, the energy barriers for the 90 ° switching and the 180 ° switching (reversing rotation) are quite similar, which supports the hypothesis that both 90 ° and 180 ° polarization flipping happen in $Sr_3Sn_2O_7$ under an external electric field. This explains why the remnant polarization along [100] is larger than that along [110] as shown in **Figure 1d**, as well as the observed small coercivity of erasing and generating FA domains. The antipolar switching path, which has the lowest energy barrier, describes polarization flipping via motion of "stacked" domain walls along the $c$ (long) axis. This type of switching has been observed in polycrystalline $Sr_3Sn_2O_7$ samples.[18]

**2.4 Creation of charged and neutral ferroelectric domain walls with strain**

As discussed above, the straight orthorhombic twin boundaries are easily idendified from t-POM. To map the four in-plane polarization directions, we use in-plane piezo-response force microscope (IP-PFM) measursments. Taking the red rectangular area shown in **Figure 2e** as an example, **Figure 5a** and **5b** present the IP-PFM images of that area, where the cantilever is rotated by 90° on going from **Figure 5a** to **5b**. Since only polarization perpendicular to the scanning cantilever can be detected, the corresponding polarization direction represented by the white (black) arrows can be identified. **Figure 5c** shows the distribution of ferroelectric (FE) domains obtained from combining the horizontal and vertical IP-PFM images. As shown in **Figure 6**, we then systematically record the change of FA and FE domains under optical microscopy and piezoresponse force microscopy when a uniaxial stress is applied to generate and expand one orthorhombic twin region. **Figure 6a** and the corresponding IP-PFM image shown in **Figure 6e** reveal the initial abundant charged domain walls with the preferred up polarization direction. In the newly created FA region B shown in Figure **6b-6c** and **6f-6g**, the FE domains with left or right polarization are both possible. This can be understood as the energy barriers for ±90° switching are the same in terms of uniaxial stress. Still, we notice one preferred polarized direction in the induced region.

The domain wall conductivity can be sharply increased if the bound charge is discontinuous at the wall, which creates a strong electric field and free charge accumulation across the wall. The bound charge density on each wall is determined by the angle between the polarizations in the adjacent FA/FE domains and also the normalized vector direction of the wall. There are four kinds of charged FA DWs, head-to-head, tail-to-tail, head-to-tail with the anticlockwise rotation and head-to-tail with the clockwise rotation, as shown in the cartoon at the top **Figure 6d**. As a result, the charge density on the smoothly curved boundaries surrounding the two island FE domains in **Figure 6g** will vary continuously from $2P$ (red) to $-2P$ (green) as shown in the bottom schematic picture of **Figure 6d**. The charge density on each FA and FE DW extracted from **Figure 6g** is shown in **Figure 6h**. A portion of the created FA domain walls are charged, although they are mostly neutral. While some of the FE domain walls in the created region B have non-zero bound charge, control over the charge density by uniaxial stress remains elusive. However, just being able to create new FA/FE domain walls provides the exciting possibility to manufacture conducting charged domain walls.

## 3. Discussion

Since ferroelectric switching in HIFs requires the reversal of an octahedral rotation, it is plausible that the switching barrier correlates with the octahedral rotation amplitude. This would suggest that the HIFs with the smallest switching barriers (and coercive fields) are those with the smallest octahedral rotation amplitudes. To explore this idea further, in Table 1 we report the octahedral tilting and rotation amplitudes calculated from DFT and obtained from experimentally reported structures for several $A_3B_2O_7$ materials and the reported polar to non-polar $T_c$. As expected, the octahedral rotation amplitudes correlate with the tolerance factor of the corresponding $ABO_3$ perovskite so that materials with smaller tolerance factors exhibit larger octahedral rotation amplitudes (and polar distortions, due to the trilinear coupling between octahedral rotations and polarization in HIFs). The material with the tolerance factor closest to 1 is $Ca_3Mn_2O_7$. However, no meaningful polarization measurment can be obtained at room temperature due to leakiness of the sample.[22] Thus, $Sr_3Sn_2O_7$ is the experimentally switchable HIF with tolerance factor closest to 1, and thus correspondingly the smallest octahedral rotation amplitudes.

Interestingly, upon contrasting the $a^-a^-c^0$ octahedral tilt and $a^0a^0c^+$ rotation amplitudes in Table 1 for $Ca_3Ti_2O_7$ and $Sr_3Sn_2O_7$ (which are both experimentally switchable HIFs) we notice that the tilt amplitudes in the two materials are almost identical, while the rotation amplitude in $Sr_3Sn_2O_7$ is significantly smaller than that in $Ca_3Ti_2O_7$. This suggests that the amplitude of the $a^0a^0c^+$ octahedral rotation, in particular, may be key for determining the energy barrier for polarization switching. To further support this idea, we note that all the low energy barrier switching paths in **Figure 4** (the one-step rotation reversal and the two-step 90º switching and antipolar paths) involve turning the $a^0a^0c^+$ rotation off and on again. In the one-step path, the $a^0a^0c^+$ rotation reverses in all n=2 perovskite slabs, while in the two-step 90º switching and antipolar paths, the $a^0a^0c^+$ rotation sense reverses in alternating perovskite slabs in each step. The $a^-a^-c^0$ octahedral tilt axis also rotates by 90° along the 90° switching path, but this occurs while maintaining an approximately constant octahedral tilt amplitude. This analysis suggests that searching for HIFs with small $a^0a^0c^+$ octahedral rotation amplitudes may be a pathway to realizing even lower coercive fields. However, since the polarization amplitude is generally linked to the rotation amplitude, a challenge is to identify a strategy to maintain a robust polarization while reducing the octahedral rotation. Note that the coexistence of the tilt and rotation modes is the prerequisite to form a polar phase, and those with a fully suppressed rotation are no longer good candidates for HIFs[3][31]. Moreover, increasing the tolerance factor and decreasing the rotation amplitudes inevitably decreases the ferroelectric $T_c$, as shown in Table 1. The $T_c$ of $Sr_3Sn_2O_7$ is 410 K and close to room temperature, at which experiments were performed. Thus, the low $T_c$ naturally goes together with the low switching barrier. Finally, we note that a small coercive stress has been reported in $Ca_3Ru_2O_7$ [19]. As shown in Table 1, due to its relatively small tolerance factor $Ca_3Ru_2O_7$ has a much larger rotation amplitude than $Sr_3Sn_2O_7$, so it may be an exception to the low barrier - small rotation amplitude trend. Note that $Ca_3Ru_2O_7$ is the only metallic compound in Table 1, which could play a possible role.

## 4. Conclusion

We report the first successful growth of a $Sr_3Sn_2O_7$ single crystal using a laser floating zone furnace and find the smallest polarization switching coercive field (39 kV cm$^{-1}$ along [100] and 47 kV cm$^{-1}$ along [110]) at room temperature among all discovered bulk hybrid improper ferroelectrics. This small coercive field can be explained from three factors: first, the single-crystallinity prevents the pinning effect and leakages which arise from grain boundaries in polycrstalline samples, second, the small energy barriers for the 90° and antipolar polarization switching paths and third, the tolerance factor being closest to 1 and correspondingly the smallest octahedral rotation amplitudes among all experimentally switchable hybrid improper ferroelectrics. Due to the small switching barrier, we observe the erasing and re-writing of ferroelastic domains with strain. The strain can erase domains with polarization parallel to it and re-write domains with polarization perpendicular to it since the shorter *b*-axis is preferred to be aligned along the strain while the polarization is along the longer *a*-axis. Moreover, we determine the ferroelectric polarization direction in each ferroelectric/ferroelastic domain from in-plane piezo-response force microscope images. We also present the bound charge density map for the four types of FA DWs and observe a continuously changing charge density varying from *2P* to *-2P* at FE DWs surrounding FE "island" domains. Our discoveries reveal the smallest coercive field among all bulk HIFs and the reproducible erasing (writing) of FE domains and charged FA (FE) domain walls, which provides new insights and application opportunities of HIFs in functional electronic devices.

## 5. Experimental Section

Material synthesis: High-purity $SrCO_3$ and $SnO_2$ powder in molar ratio 1:1 were mixed thoroughly, calcined at 1200 ℃ for 10 hours and sintered at 1500 ℃ for 10 hours. The derived powder was pelletized into a uniform rod shape by 8000 PSI hydrostatic pressure and sintered at 1600 ℃ for 10 hours to get a high-density feed rod. Note that the 50% excessive $SnO_2$ in the feed rod is to compensate the evaporation loss during growth. The single crystal was grown at the speed of 30 mm per hour in a 0.9 MPa $O_2$ atmosphere. High oxygen pressure helps to suppress $SnO_2$ evaporation and stabilize the $Sr_3Sn_2O_7$ phase. The as-grown $Sr_3Sn_2O_7$ single crystal boule, as shown in **Figure 1a**, was annealed at 1400 °C for 20 hours followed by a 50 °C per hour cooling in $O_2$ flow to diminish defects and oxygen vacancies.

PFM measurements: For all the PFM measurements, AC 5 V at 44 kHz was applied to a tip (a commercial conductive AFM tip) while sample backside is grounded.

PE measurement: The PE loops were measured on an ab-plane cleaved thin crystal by the "PUND" method provide in the Ferroelectric Material Test System (RADIANT TECHNOLOGIES INC.). The remnant-only polarization hysteresis loop is finally derived.

DFT calculations: We perform DFT calculations using the Vienna Ab Initio Simulation Package (VASP)[32] [33] with the PBEsol[34] exchange correlation functional. We use a 600 eV plane-wave energy cutoff and a 6 x 6 x 2 Monkhorst-Pack k-point mesh, where the computational cell contained 48 atoms and could accommodate both orthorhombic twins. For structural relaxations, we use a force convergence tolerance of 2 meV $A^{-1}$. For the $Ca_3Mn_2O_7$ calculation reported in Table 1, we used DFT+U with U=4.5 eV and J = 1 eV. The energy barriers for the one-step and two-step 90 ° ferroelectric switching paths are obtained from bulk structural relaxations with the symmetry constrained to that of the barrier structure: *Acam* for the one-step path along which the $a^-a^-c^0$ tilt reverses, *Amam* for the one-step path along which the $a^0a^0c^+$ rotation reverses, and *C2mm* for the two-step 90 ° switching path. The intermediate points along the ferroelectric switching paths between the ground state $A2_1am$ structure and the barrier structure are obtained from nudged elastic band (NEB) calculations, where the atomic positions are allowed to relax for each image along the path. For the two-step antipolar switching path, we obtain the energy barrier from our NEB calculations that connect the $A2_1am$ structure to the antipolar (symmetry *Pnam*) structure. This is because all the intermediate points between $A2_1am$ and *Pnam* have the same symmetry, $P2_1am,$ so we cannot calculate the barrier energy via structural relaxation in this case. These calculations yield intrinsic single-domain ferroelectric switching paths, so the energy barriers correspond to the energy it takes to reverse the polarization in a single infinite domain and can be viewed as upper limits for barriers for real switching, mediated by domain wall motion and domain nucleation. We perform group theoretic analysis with the ISOTROPY Software Suite.[35] The structural distortion amplitudes reported in Table 1 are obtained by decomposing the distorted $A2_1am$ structure with respect to the high-symmetry $I4/mmm$ reference structure. The $a^-a^-c^0$ octahedral tilt, $a^0a^0c^+$ octahedral rotation, and polar distortion transform like the $X_3^-$, $X_2^+$, and $\Gamma_5^-$ irreducible representations of $I4/mmm$, respectively. The lattice constants calculated from DFT are shown in Table 2.

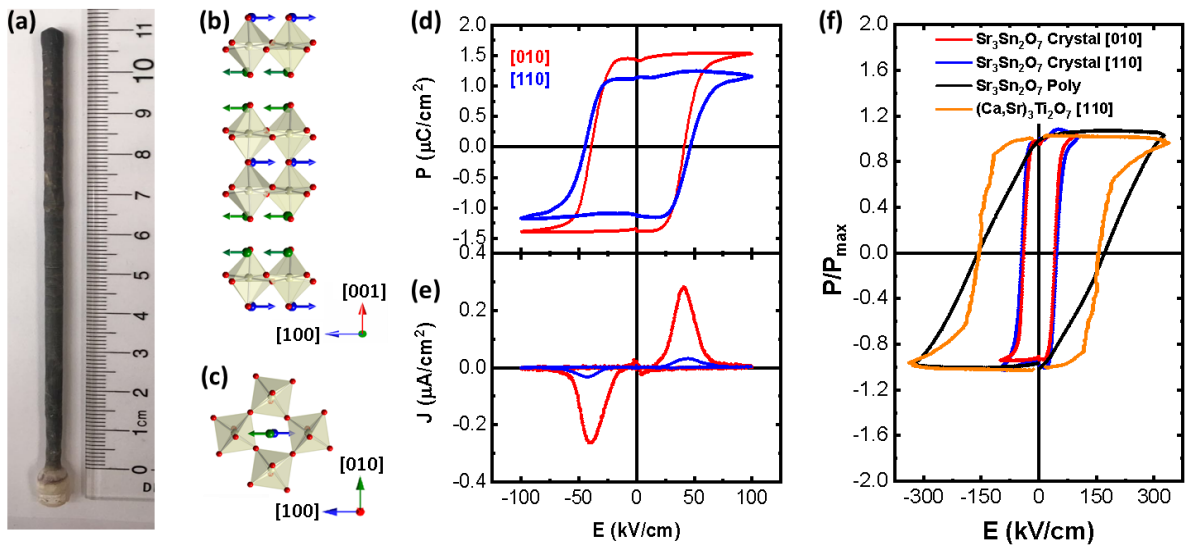

**Figure 1. Switchable electric polarization at room temperature with smallest coercive field among all HIF bulk materials.**

(a) Photograph of an as-grown $Sr_3Sn_2O_7$ single crystal boule. (b, c) Side-view and top-plane view of the $A2_1am$ structure of $Sr_3Sn_2O_7$ consisting of an $a^-a^-c^+$ octahedral rotation pattern and polar displacements. The blue and green arrows represent the Sr displacements along the *a*-axis in the perovskite block and rocksalt block, respectively. The red (grey) dots depict oxygen (Sn) atoms. (d, e) Electric polarization (*P*) and compensated current density (*J*) versus electric field (*E*) hysteresis loops performed on a $Sr_3Sn_2O_7$ single crystal, measured along two different crystallographic directions with a PUND method at room temperature at frequency *f*=270 Hz. The red (blue) curves represent the data measured with the field along the [010] ([110]) direction. (f) The normalized *P(E)* loops of a $Sr_3Sn_2O_7$ single crystal (blue along [110] and red along [100]), $Ca_3Ti_2O_7$ single crystal (yellow, along [110]) and $Sr_3Sn_2O_7$ poly sample (black).

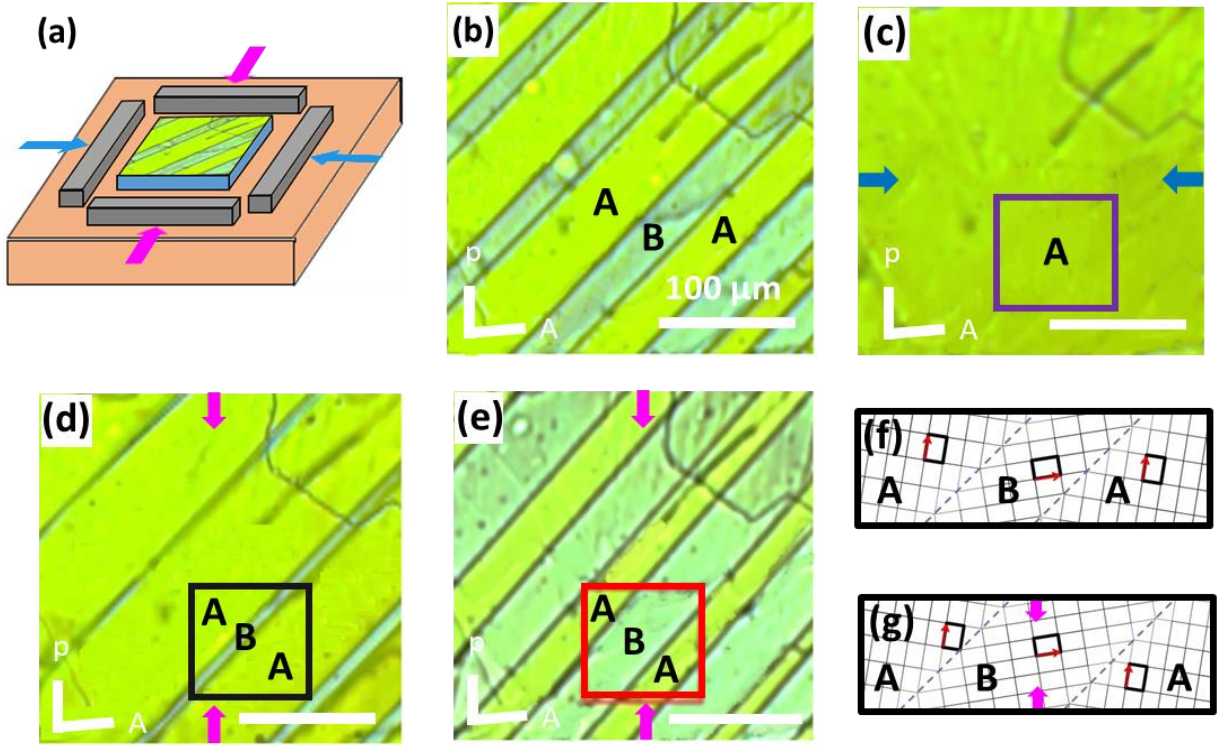

**Figure 2. Erasing (shrinking) and re-generating (expanding) orthorhombic twin domains in a $Sr_3Sn_2O_7$ single crystal.**

(a) A schematic geometry of the experiment. An *ab*-plane cleaved thin piece is polished into a square shape with boundaries along the crystal [100] ([010]) directions. (b-e) Transmission polarized optical microscope (t-POM) images. The dark-bright contrast in (b, d, e) indicates the orthorhombic twin domains. (b) is the initial state and (c-e) present the evolution of twin domains under uniaxial stress applied in two directions, as shown by the blue arrows (horizontal) and pink arrows (vertical) in each figure. The colored squares in (c-e) indicate the corresponding scanned regions, whose PFM images are shown in Figure 6. In (b-e), the white bar marked by P (A) represents the direction of the polarizer (analyzer). (f-g) A cartoon showing how FA domains change under stress. The bold rectangle represents an *ab*-projection of the orthorhombic unit cell. The red arrow indicates the polarization direction, which lies along the crystallographic *a* axis. The dashed lines denote the FA domain walls. (f) Shows the initial state and (g) shows the final state after a compressive stress has been applied in the vertical direction as indicated by the thick pink arrows.

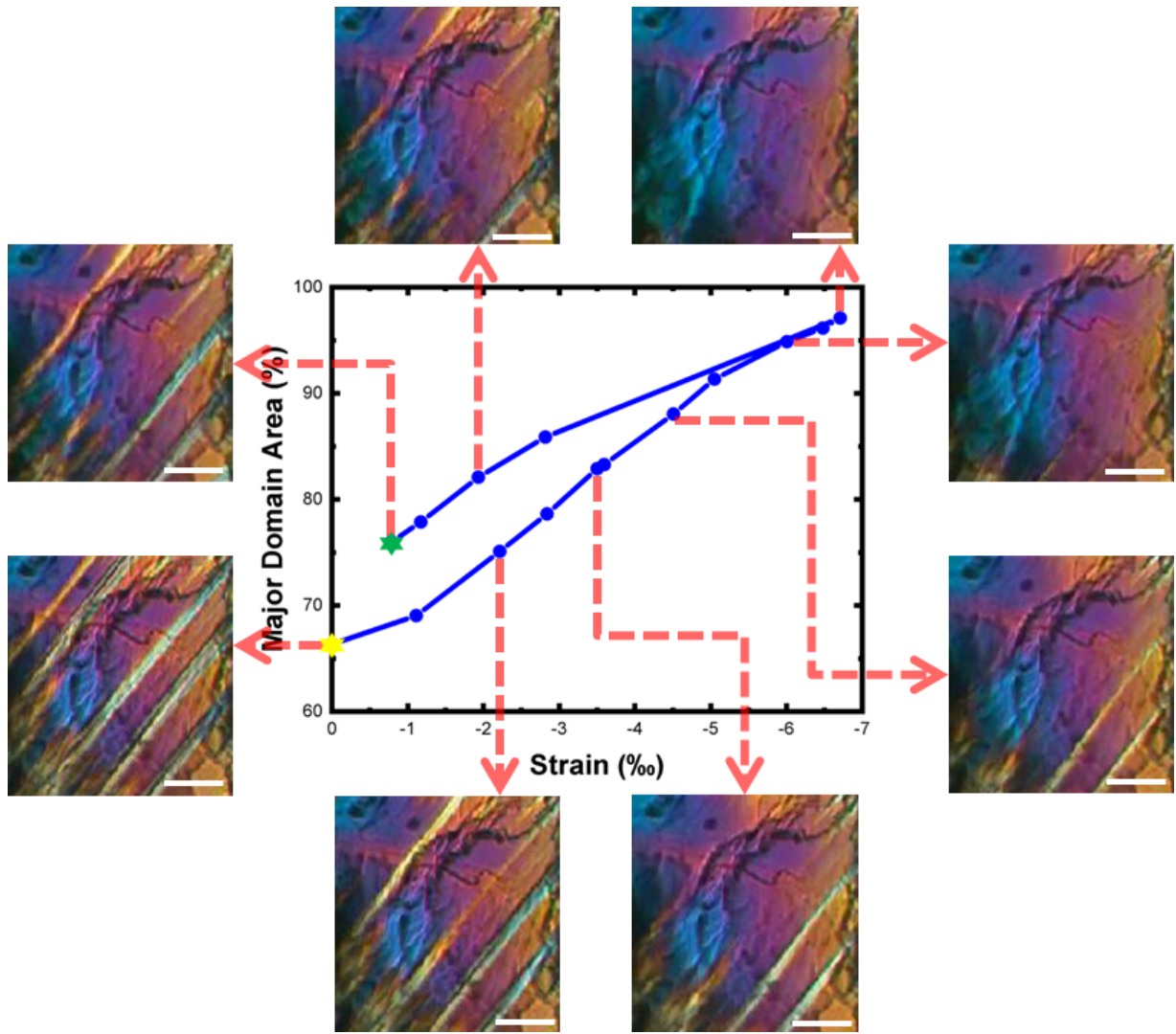

**Figure 3. Strain hysteresis half loop.**

A strain hysteresis half loop derived from domain area statistics in a stress-and-release process. Each t-POM photo reflects the domain distribution which produces the data point indicating by the red dashed arrow. The yellow and green stars represent zero-stress states right before and after the test, respectively. The white scale bar is 50 μm for each photo.

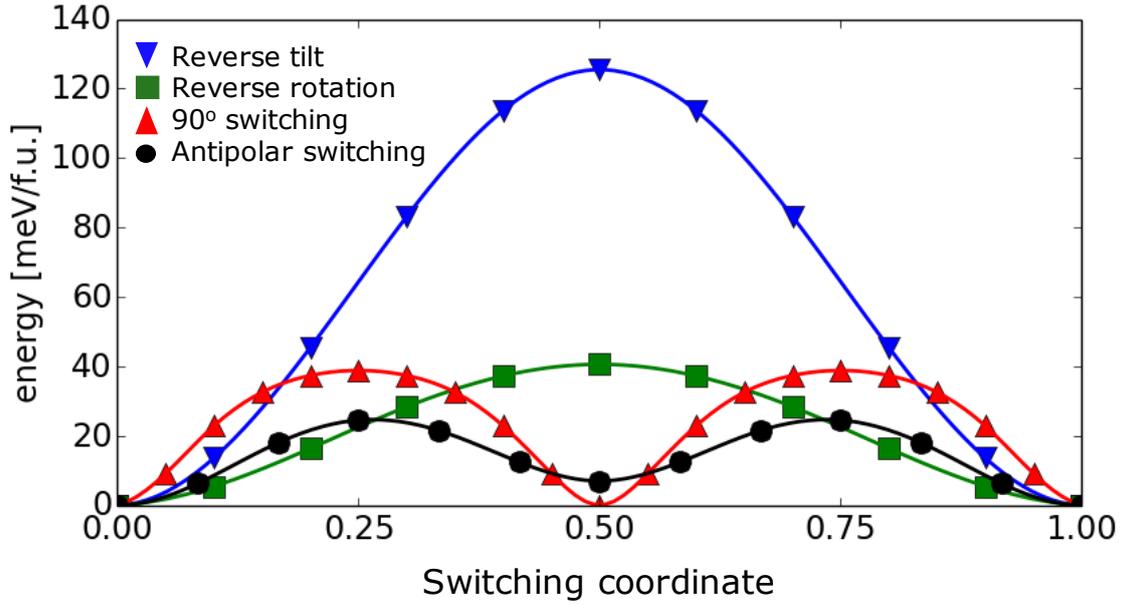

**Figure 4. Ferroelectric switching paths for $Sr_3Sn_2O_7$ calculated with DFT.**

The energy barriers for the paths are 125 meV f.u.$^{-1}$ (reverse tilt), 41 meV f.u.$^{-1}$ (reverse rotation), 39 meV f.u.$^{-1}$ (two-step 90° switching), and 24 meV f.u.$^{-1}$ (two-step switching via antipolar phase).

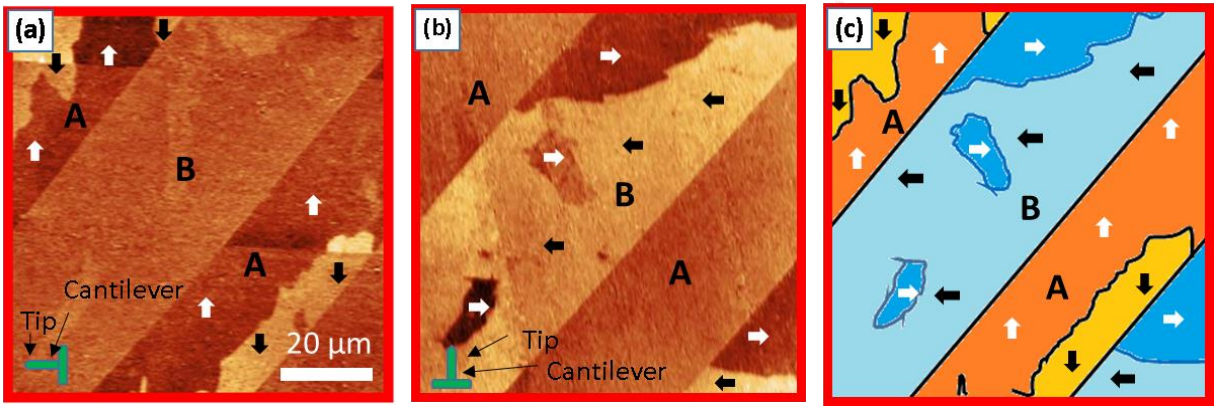

**Figure 5. Polarization direction in each ferroelectric domain.**

(a-b) In-plane piezo-response force microscope (IP-PFM) images of the region shown in Figure 2e. Orthorhombic twin boundaries are oriented along the diagonal direction of the *xy* scanning axes. In (a) and (b), the long axis of the AFM cantilever is oriented along the horizontal and vertical directions respectively, as shown by the cartoon in the left bottom corner. (c) Illustration of polar domains in an 80 μm × 80 μm area, obtained from (a) and (b).

The black and white arrows in (a-c) indicate the direction of the electric polarization in each domain.

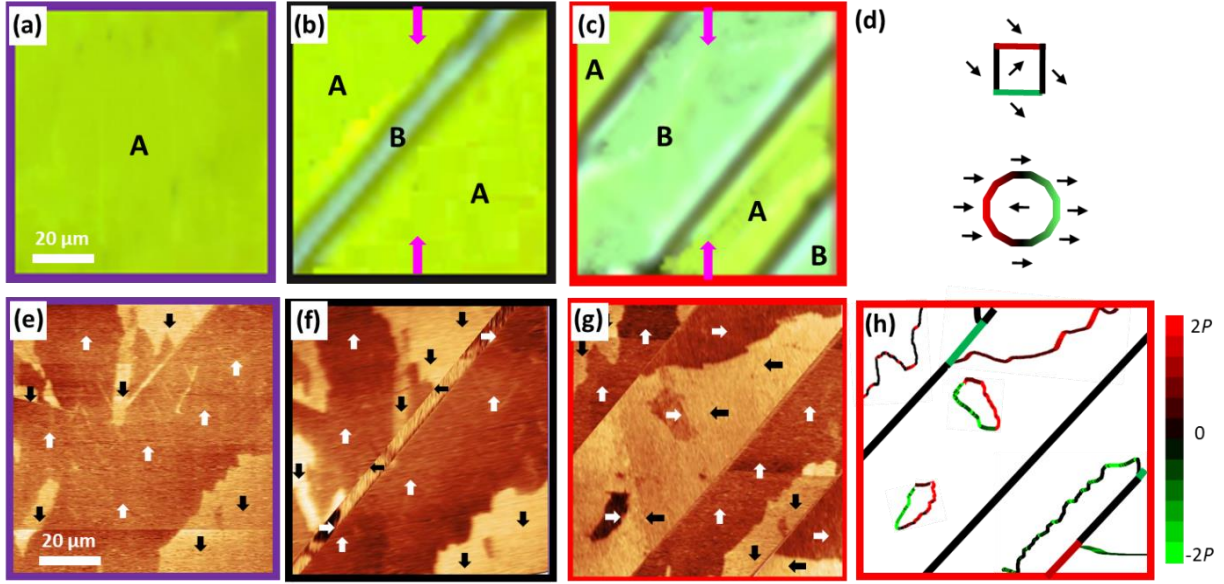

**Figure 6. Effect of stress on ferroelectric-ferroelastic domains and charged domain walls.** (a-c) T-POM images of the region marked by a square in Figure 2c-2e. (d) For the FA DWs, there are four different DWs with three different charge densities (0, $\pm\sqrt{2}P$) as shown in the top of panel (d). For the FE DWs that surround the two island FE domains in (g), the charge density changes continuously from $2P$ to $-2P$. The bottom part of panel (d) depicts the possible charge densities at FE DWs with respect to the polarization direction (black arrows). The colors in (d) match the color bar given in (h), which represents the variation of charge denstiy from $2P$ to $-2P$. (e-g) The corresponding IP-PFM images with white and black arrows indicating the polarization direction in each domain. (h) A schematic depicting the bound charge density of all the FA and FE DWs shown in (g).
17

**Table 1. Tolerance factors, structural distortion amplitudes, and $T_c$ of $A_3B_2O_7$ materials with symmetry $A2_1am$.**

| Material | $t$ | $Q_T$ [DFT] | $Q_R$ [DFT] | $Q_T$ [Exp] | $Q_R$ [Exp] | $T_c$ [Exp] | Exp. Ref. |
|---|---|---|---|---|---|---|---|
| $Ca_3Mn_2O_7$ | 0.977 | 0.87 | 0.79 | 0.66 | 0.73 | 270-320 K | [22][36] |
| $Sr_3Sn_2O_7$ | 0.957 | 1.18 | 0.76 | 0.89 | 0.67 | 410 K | [18] |
| $Ca_3Ti_2O_7$ | 0.946 | 1.20 | 0.87 | 1.01 | 0.83 | 1100 K | [21][37] |
| $Sr_3Zr_2O_7$ | 0.942 | 1.27 | 0.86 | 1.07 | 0.78 | 700 K | [6] |
| $Ca_3Ru_2O_7$ | 0.937 | 1.40 | 1.11 | 1.38 | 1.10 | >1073 K | [38] |
| $Ca_3Sn_2O_7$ | 0.905 | 1.88 | 1.17 | - | - | - | - |
| $Ca_3Zr_2O_7$ | 0.891 | 1.94 | 1.22 | - | - | - | - |

Perovskite tolerance factor $t$, $Q_T$[DFT] of the $a^-a^-c^0$ octahedral tilt and $Q_R$[DFT] of the $a^0a^0c^+$ octahedral rotation from DFT calculation, $Q_T$[Exp] of the $a^-a^-c^0$ octahedral tilt and $Q_R$[Exp] of the $a^0a^0c^+$ octahedral rotation from experimentally determined structure, and reported $T_c$ for several $A_3B_2O_7$ materials with symmetry $A2_1am$. The DFT distortion amplitudes are reported for a 24 atom cell, and the distortion amplitudes and lattice parameters are given in Å.

**Table 2. DFT calculated lattice constants of $A_3B_2O_7$ materials with symmetry $A2_1am$.**

| Material | $a$(Å) | $b$(Å) | $c$(Å) |
|---|---|---|---|
| $Ca_3Mn_2O_7$ | 5.22 | 5.22 | 19.23 |
| $Sr_3Sn_2O_7$ | 5.76 | 5.74 | 20.62 |
| $Ca_3Ti_2O_7$ | 5.44 | 5.39 | 19.30 |
| $Sr_3Zr_2O_7$ | 5.81 | 5.80 | 20.80 |
| $Ca_3Ru_2O_7$ | 5.47 | 5.28 | 19.62 |
| $Ca_3Sn_2O_7$ | 5.73 | 5.56 | 19.61 |
| $Ca_3Zr_2O_7$ | 5.78 | 5.61 | 19.72 |